\pdfoutput=1

\documentclass[11pt,a4paper]{article}
\usepackage{jcappub}
\usepackage{amsmath}
\usepackage{bm}
\usepackage{color}
\usepackage{subfig}
\usepackage{comment}
\usepackage{graphicx}
\usepackage[style=numeric-comp,sorting=none,backend=bibtex]{biblatex}
\def\bea{\begin{eqnarray}}
\def\eea{\end{eqnarray}}

\title{Neutrino point source searches for dark matter spikes}

\author[a,b,c]{Katherine Freese,}

\author[a]{Irina Galstyan,}

\author[d]{Pearl Sandick,}

\author[a,e,f,g]{Patrick Stengel}

\affiliation[a]{The Oskar Klein Centre for Cosmoparticle Physics,
	Department of Physics,
	Stockholm University,
	AlbaNova,
	10691 Stockholm,
	Sweden}
\affiliation[b]{Nordita,
	KTH Royal Institute of Technology and Stockholm University
	Roslagstullsbacken 23,
	10691 Stockholm,
	Sweden}
\affiliation[c]{
    Department of Physics,
  University of Texas,
  Austin, TX 78722}
\affiliation[d]{Department of Physics and Astronomy, University of Utah, Salt Lake City, UT 84102, USA}
\affiliation[e]{Scuola Internazionale Superiore di Studi Avanzati (SISSA), via Bonomea 265, 34136 Trieste, Italy}
\affiliation[f]{INFN, Sezione di Trieste, via Valerio 2, 34127 Trieste, Italy}
\affiliation[g]{Institute for Fundamental Physics of the Universe (IFPU), via Beirut 2, 34151 Trieste, Italy}

\abstract{
Any dark matter spikes surrounding black holes in our Galaxy are sites of significant dark matter annihilation, leading to a potentially detectable neutrino signal. In this paper we examine $10-10^5 M_\odot$ black holes associated with dark matter spikes that formed in early minihalos and still exist in our Milky Way Galaxy today, in light of neutrino data from the ANTARES~\cite{ANTARES:2017dda} and IceCube~\cite{IceCube:2019cia} detectors. In various regions of the sky, we determine the minimum distance away from the solar system that a dark matter spike must be in order to have not been detected as a neutrino point source for a variety of representative dark matter annihilation channels. Given these constraints on the distribution of dark matter spikes in the Galaxy, we place significant limits on the formation of the first generation of stars in early minihalos---stronger than previous limits from gamma-ray searches in Fermi Gamma-Ray Space Telescope data. The larger black holes considered in this paper may arise as the remnants of Dark Stars after the dark matter fuel is exhausted; thus neutrino observations may be used to constrain the properties of Dark Stars. The limits are particularly strong for heavier WIMPs. For WIMP masses $\sim 5 \,$TeV, we show that $\lesssim 10 \%$ of minihalos can host first stars that collapse into BHs larger than $10^3 M_\odot$.
}

\notoc
\makeatletter
\gdef\@fpheader{\begin{flushright}
NORDITA-2022-002; UTTG-32-2022
\end{flushright}}
\makeatother

\begin{document}

\maketitle

\section{Introduction}

The nature of dark matter (DM) is one of the most interesting open questions in particle physics, and the discovery of a DM signal could offer the first hints of physics beyond the Standard Model (SM). While the existence of DM has been well established by its gravitational interactions at a variety of scales over several different epochs, the detection of DM interactions with the SM have remained elusive in an increasingly broad experimental program. One of the best-motivated DM candidates is the weakly interacting massive particle (WIMP), signatures of which can potentially be detected when produced at collider experiments or when recoiling off SM particles at direct detection experiments. In addition, the SM products of WIMP annihilation in the Milky Way halo can be searched for in terrestrial or space-based indirect detection (ID) experiments. 

The WIMP signal at ID experiments depends on both the annihilation rate and the particular composition of the SM annihilation products arising from a variety of potential annihilation channels (for a recent overview, see e.g.~\cite{Leane:2020liq}). Charged particles such as electrons, positrons and (anti)protons produced by WIMP annihilation can be detected in cosmic rays incident on Earth after propagating through the galaxy and the solar system. ID searches for gamma-rays originating from the products of WIMP annihilation can be especially sensitive since experiments such as H.E.S.S~\cite{Lefranc:2016srp} and Fermi-LAT~\cite{Fermi-LAT:2016uux} are sensitive to photons originating in parts of the galaxy with either a large potential DM signal (e.g. galactic center) or a small SM background (e.g. dwarf satellite galaxies). Neutrino detection experiments such as ANTARES (AN) and IceCube (IC) are also sensitive to WIMP annihilation in the galaxy~\cite{Albert:2016emp,IceCube:2016oqp,IceCube:2017rdn,ANTARES:2019svn,ANTARES:2020leh}, as well as to the annihilation of DM that is gravitationally captured in the Earth and Sun~\cite{Freese:1985qw,Krauss:1985aaa,Silk:1985ax,Baum:2016oow,ANTARES:2016bxz,ANTARES:2016xuh,IceCube:2016dgk,IceCube:2016aga}. While ID of both gamma-rays and neutrinos is not subject to most of the uncertainties associated with cosmic ray propagation, the relative sensitivity of the respective experiments to WIMP annihilation is largely determined by whether the WIMPs annihilate through channels that primarily yield hadronic or leptonic annihilation products.

Both N-body simulations of structure formation and observations of galaxies besides the Milky Way suggest that the DM density in the halo either strongly peaks in the galactic center or has more of a cored profile due to the effects of baryonic feedback. In both cases, the dominant contribution to the ID signal arises from the center of the smooth and isotropic halo component of the galaxy's DM distribution. While the backgrounds in ID searches for gamma-rays from dwarf satellites are expected to be negligible, and the background neutrinos in searches for DM bound in the Sun (or Earth) can be largely suppressed using directional information, the signal from WIMP annihilation in such searches is relatively small. However, localized enhancements to the WIMP annihilation rate within the galaxy are possible through a variety of mechanisms, including those that lead to the formation of significant overdensities referred to as DM spikes, as studied here. 

DM spikes can form around black holes (BHs) at various times in the history of the Universe. For example, DM spikes can begin to form in the very early universe as WIMPs accrete around primordial black holes (PBHs) during the radiation dominated epoch (for example, see~\cite{Lacki:2010zf,Eroshenko:2016yve,Nishikawa:2017chy,Boucenna:2017ghj,Adamek:2019gns,Boudaud:2021irr}). While the details of the DM spike formation can vary significantly depending on the PBH mass and evolution of the thermal plasma, WIMP annihilation within the spikes typically yields DM density profiles characterized by a maximum density close to the center of the spike which is primarily dependent on the WIMP mass and annihilation cross section. The DM in the spikes formed around PBHs that remains today is potentially detectable with ID searches. 

In this paper, we instead investigate the ID signatures of WIMPs annihilating in DM spikes formed at later times, e.g. around BHs remaining from the epoch of the formation of the first stars at $z \sim 10-50$. N-body simulations of galaxies similar to the Milky Way predict a significant number of $\sim 10^6M_\odot$ DM minihalos in which the first stars could arise.  We will use the terminology Pop III.1 to refer to the first stars regardless of their nature, i.e. whether they are standard fusion-powered stars or Dark Stars (DSs) powered by DM heating. Once the power source for the first stars runs out, whether it be fusion or DM annihilation, these Pop III.1 stars would eventually collapse to form BHs, around which the DM within the host minihalos would contract and yield a population of spikes. Similar to a DM spike formed around a PBH, the annihilation of WIMPs within a spike formed around a BH both limits the DM density at the center of the spike and produces SM products that can be detected with ID. 

While the DM spikes that formed around BHs associated with Pop III.1 stars are insensitive to the physics of the early universe, the distribution of spikes in the galaxy does significantly depend on the relatively unconstrained history of the first epoch of star formation. In particular, if the first stars form over a longer period of time, then a larger number of minihalos can potentially host them, resulting in more DM spikes contributing to the ID signal. The density profile of an individual DM spike depends on the seed BH mass, with larger DM densities contracting around heavier BHs, and also on the epoch of star formation. Standard models of stellar evolution suggest the collapse of fusion-powered Pop III.1 stars yields $10-10^2M_\odot$ BHs, but the possibility of an extended Dark Star phase during the star formation process could produce significantly more massive $10^3-10^6 M_\odot$ BHs, and lead to correspondingly larger BH spikes. 

Dark Stars (DSs) are a new type of first star, powered by DM heating rather than by fusion~\cite{Spolyar:2007qv,Freese:2015mta}.  While DSs are made almost entirely of ordinary hydrogen and helium (with only 0.01\% of the star made of DM), DM annihilation can provide their heat source. During the DS phase of stellar evolution, DM heating keeps the temperature cool enough to prolong the accretion of matter. As a consequence, a very large star, and thus a very large BH, may result. Our work, though motivated by the DS scenario, applies generally to BHs at the centers of minihalos, regardless of their origin. Many BHs of mass $10 - 10^5 M_\odot$ that formed at the centers of minihalos survive in the universe today. Assuming some fraction of high redshift minihalos hosted Population III.1 star formation, one can estimate the distribution of their remnant BHs today.

The sensitivity of ID experiments to the annihilation products of DM within a spike depends on the WIMP annihilation channel. Previous work on DM spikes formed around BHs has largely focused on constraining the earliest epoch of star formation using the gamma-ray flux measured by experiments such as Fermi-LAT~\cite{Bertone:2005xz,Bertone:2009kj,Sandick:2010yd,Sandick:2011zs}. In particular, the flux of gamma-rays from DM spikes in the galaxy could be measured either as a collection of individual point sources (PSs) or as a contribution to the diffuse flux. Due to the extensive catalogue of gamma-ray PSs which have already been identified and accounted for in models of the diffuse flux, previous analyses have required a careful separation of DM spikes which could be measured as contributions to either. Typically, the constraints on different scenarios of Pop III.1 star formation are strongest from DM spikes which are closest to the solar system and bright enough to be identified as PSs. Also, previous analyses have been most sensitive to WIMP annihilation channels with hadronic annihilation products due to the higher fraction of the available energy from the DM going into photons.

Here, we investigate the ID of neutrinos from remnant WIMP annihilation within DM spikes (for examples of previous works investigating neutrino signals from DM spikes, see~\cite{Bertone:2006nq,Bertone:2009kj,Sandick:2010yd}). Our study complements previous work focused on the ID of gamma-rays from DM spikes; neutrinos are more readily produced in leptonic annihilation channels. In general, our analysis shows that the constraints on Pop III.1  star formation from neutrino PS analyses by AN~\cite{ANTARES:2017dda} and IC~\cite{IceCube:2019cia} can be significantly more stringent than those from gamma-ray searches. The difference in sensitivity to signals from DM spikes between neutrino and gamma-ray searches is most pronounced for BH masses $\gtrsim 10^3 M_\odot$, making neutrino experiments an ideal probe of a potential DS phase in the formation of the first stars. Also, because AN and IC are better able to detect neutrinos with energies $\gtrsim 1 \,$TeV, we show that the limits on Pop III.1 star formation from WIMP annihilation to neutrinos in DM spikes are particularly strong for heavier WIMPs. For WIMP masses $\sim 5 \,$TeV across a variety of representative annihilation channels, we show that $\lesssim 10 \%$ of minihalos can host Pop III.1 stars that collapse into BHs larger than $10^3 M_\odot$.

The rest of this paper is outlined as follows. We describe the distribution of DM spikes in the galaxy and density profiles of individual spikes in Sec.~\ref{sec:spikes}. In Sec.~\ref{sec:signal}, for a variety of WIMP masses and annihilation channels, we calculate the minimum distance away from the solar system in various regions of the sky that a single DM spike must be located to not have been detected as a PS by AN or IC. The corresponding constraints on Pop III.1 star formation are presented in Sec.~\ref{sec:constraints} and we briefly summarize our results in Sec.~\ref{sec:Con}.

\section{Dark matter spikes} \label{sec:spikes}

In order to estimate the distribution of minihalos capable of hosting first stars that would have merged into the Milky Way, we use previous results of the Via Lactea-II N-body simulation. Following the parameterization of Ref.~\cite{Trenti:2009}, we assume that every minihalo with a mass
\begin{equation} \label{eqn:mmin}
M^{\rm halo} \gtrsim 1.54 \times 10^5 M_\odot \left( \frac{1+z}{31} \right)^{-2.074}
\end{equation}
may host a Pop III.1 star. Since hierarchical structure formation yields larger numbers of less massive subhalos than those which are more massive, our analysis is not particularly sensitive to the largest subhalo mass we consider for Pop III.1 star formation, which we take to be $M^{\rm halo} \simeq 10^7 M_\odot$~\cite{Ahn:2006qu}. Thus, the number of Pop III.1 stars largely depends on the epoch during which such stars can form, specifically on the redshift of the end of Pop III.1 star formation $z_f$. As this redshift is poorly constrained, we consider three different scenarios for the end of Pop III.1 star formation: Early ($z_f$ = 20), Intermediate ($z_f$ = 15), and Late ($z_f$ = 10). The number density of minihalos suitable for Pop III.1 star formation, and thus for hosting DM spikes, $N_{\rm sp}^{(z_f)}$, is defined such that the total number of minihalos is given by
\begin{equation}
    \int d^3 \mathbf{R} \, N_{\rm sp}^{(z_f)}(R) \, ,
    \label{eq:Nsp}
\end{equation}
where $R$ is the radial distance from galactic center. For the Early, Intermediate, and Late $z_f$ scenarios, the respective numbers of minihalos are 409, 7983, and 12416. In this section and the next, we assume that a Pop III.1 star forms in every suitable minihalo and subsequently collapses into a BH, with an accompanying DM spike around the BH. In Sec.~\ref{sec:constraints}, we consider the constraints from neutrino PS searches to set limits on the fraction of minihalos which yield spikes capable of producing neutrinos from DM annihilation. 

\begin{figure}
\begin{center}
\includegraphics[width=3in]{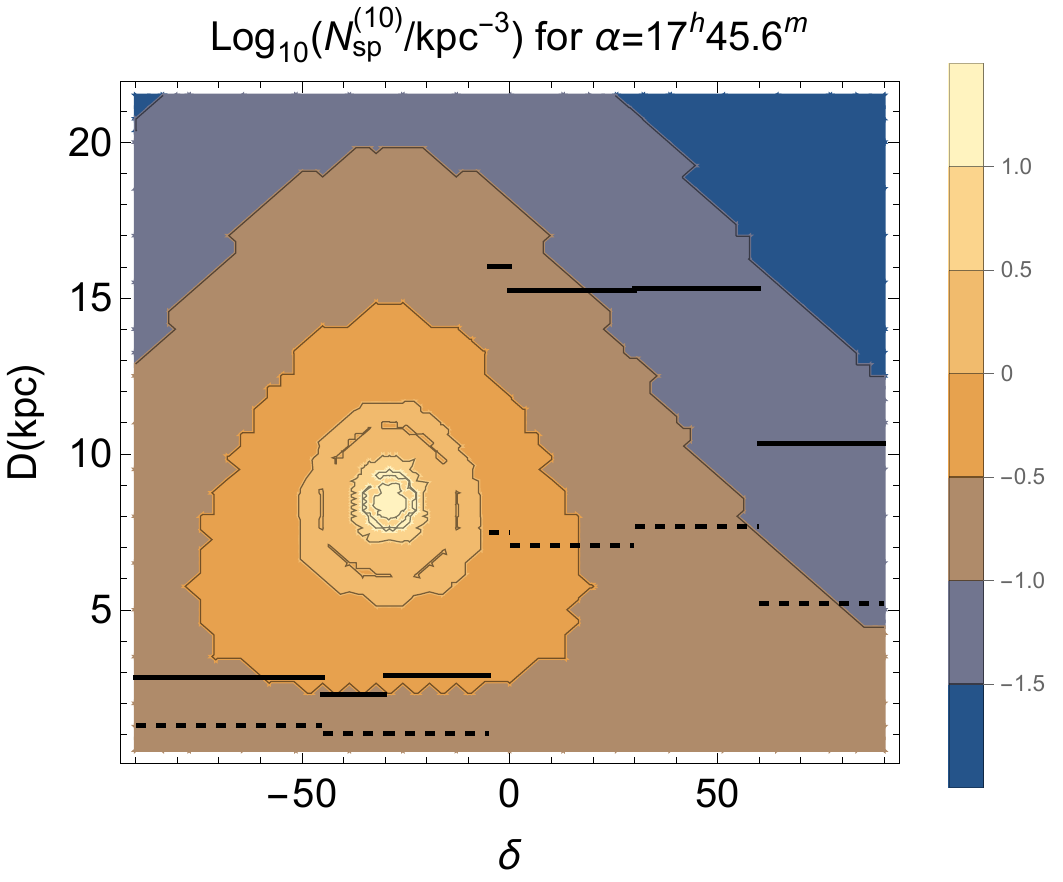}\hfill
\includegraphics[width=3in]{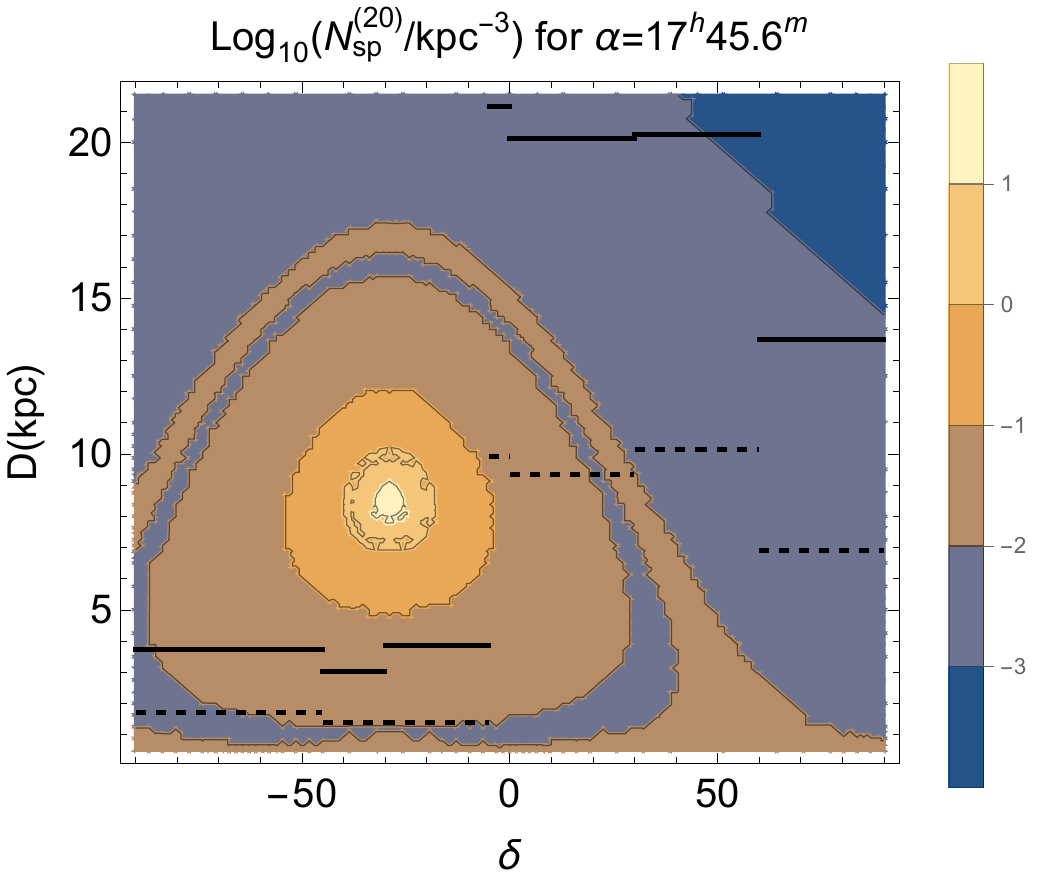}
\includegraphics[width=3in]{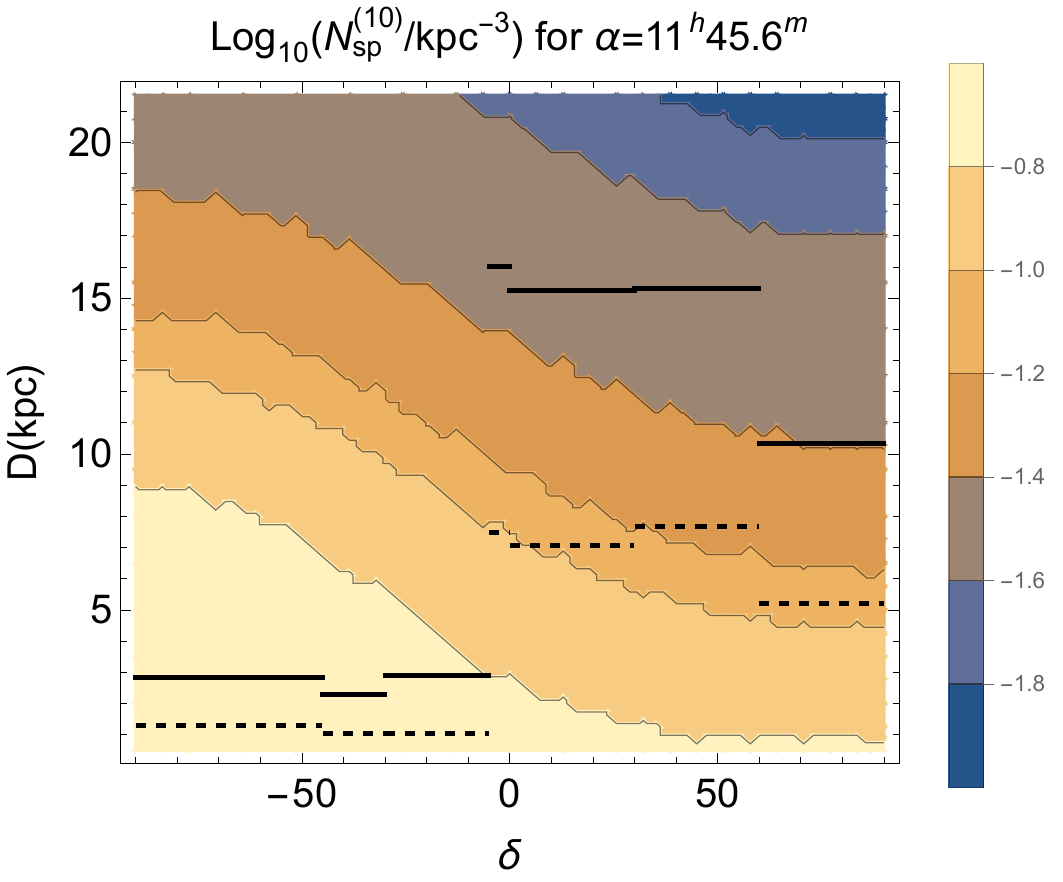}\hfill
\includegraphics[width=3in]{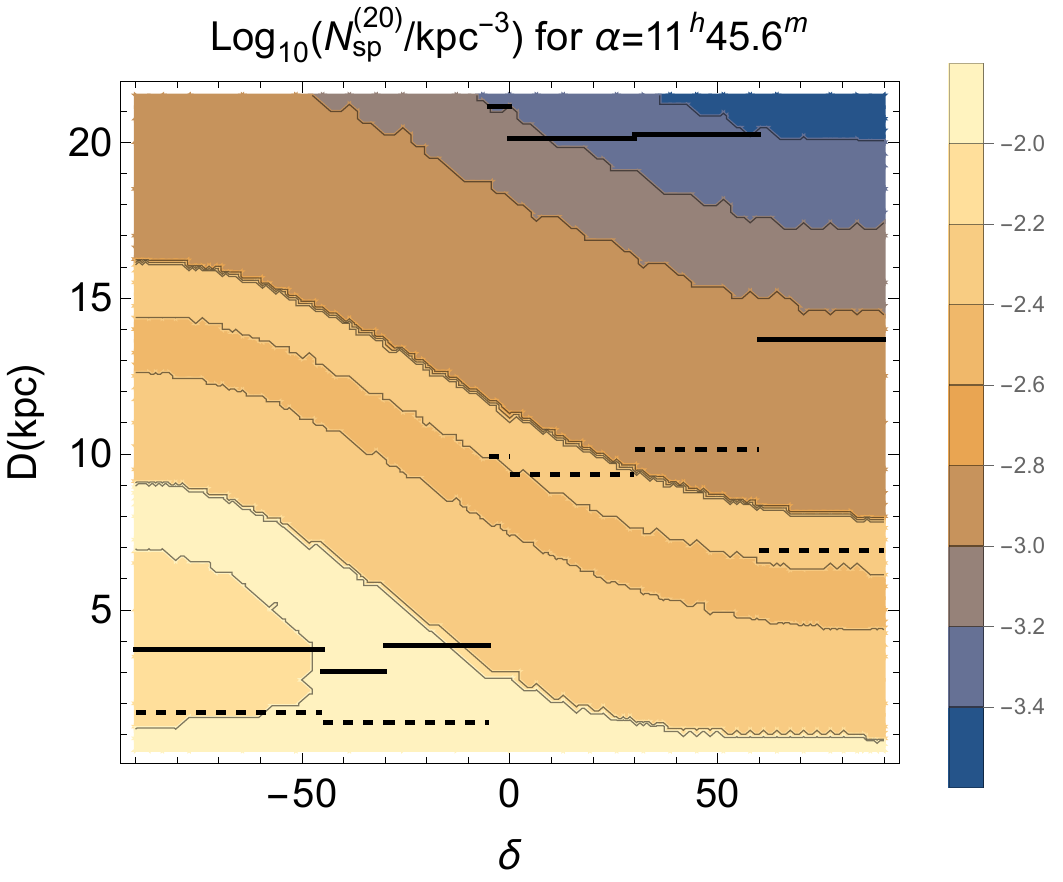}
\includegraphics[width=3in]{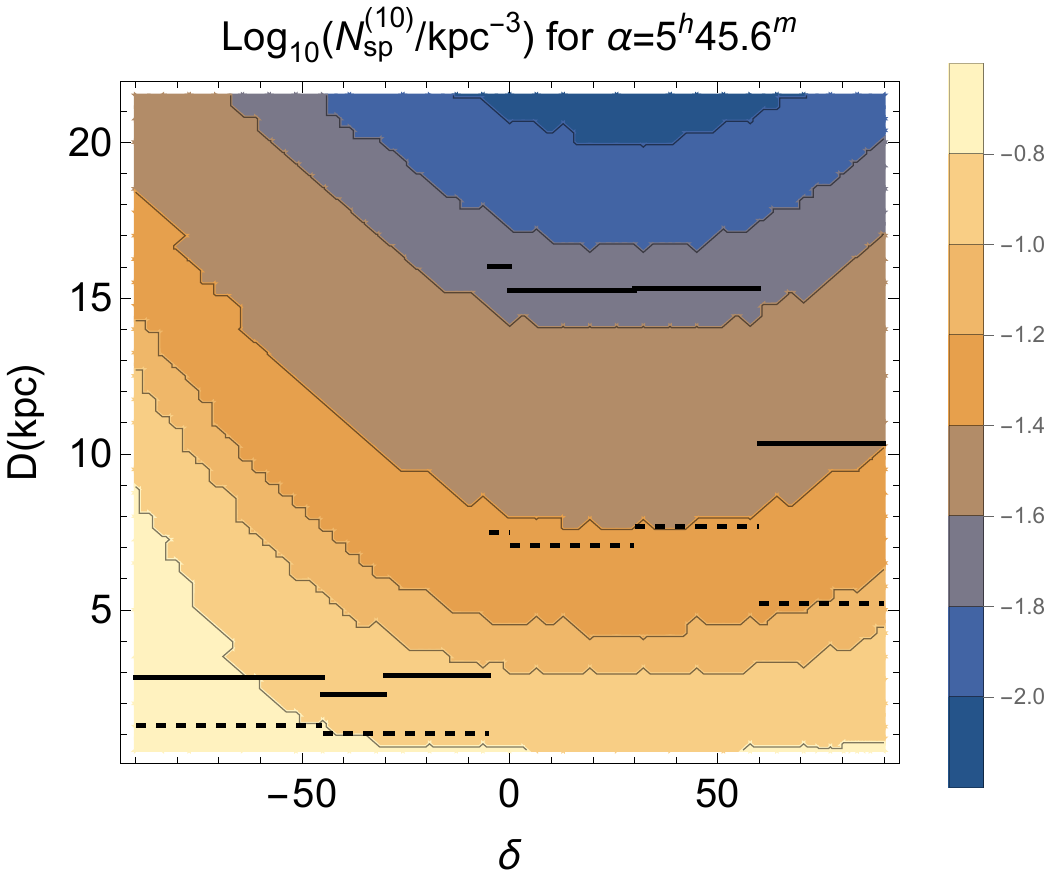}\hfill
\includegraphics[width=3in]{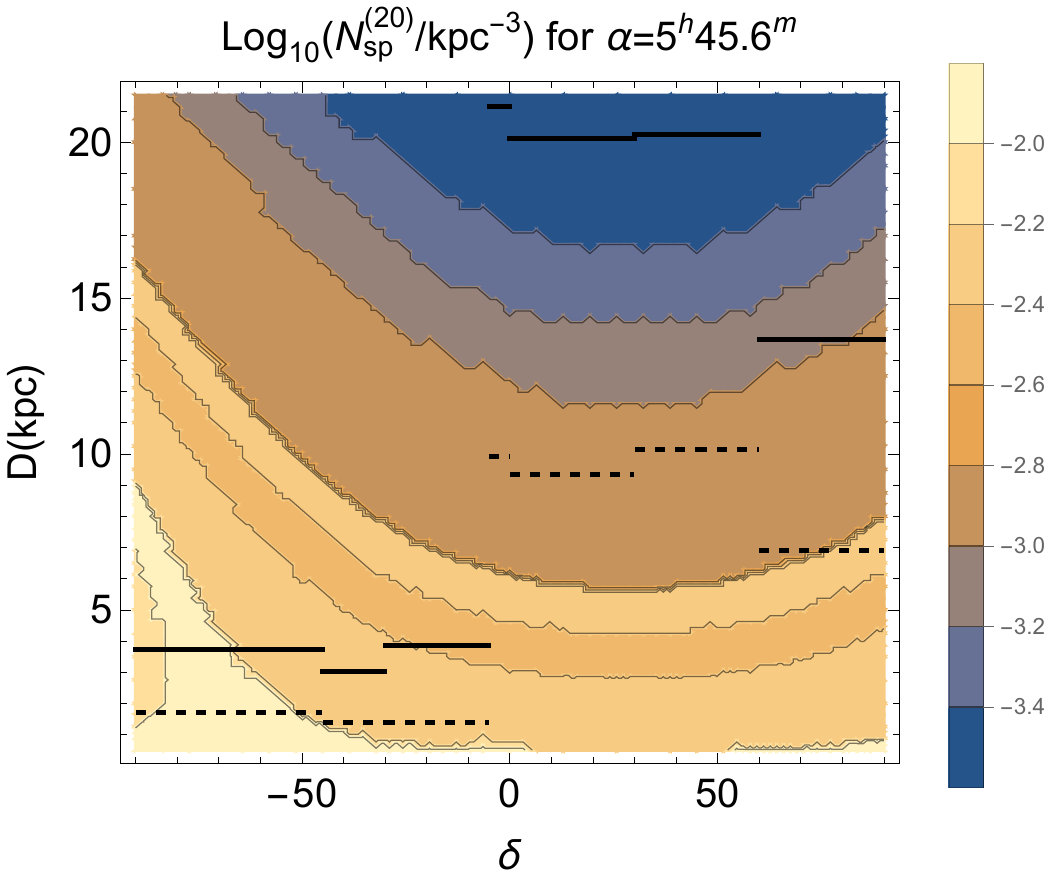}
\vspace{-0.0cm}
\caption{Contours of the density of dark matter spikes in the Milky Way, $N_{\rm sp}^{(z_f)}$, assuming the Late (left) and Early (right) termination of Pop III.1 star formation as a function of distance from the solar system, $D$, and declination angle, $\delta$. Right ascension, $\alpha$, is fixed in the direction of (top), perpendicular to (middle) and away from (bottom) galactic center. For an illustration of the analysis in Sec.~\ref{sec:signal}, we also show the minimum distance a DM spike formed around a $10^3 M_\odot$ BH must be located from the solar system, $D_{\rm min}$, in order to have not been detected in neutrino point source searches by ANTARES and IceCube, assuming a $m_\chi = 500 \,$GeV (dashed lines) and $m_\chi = 5 \,$TeV (solid lines) WIMP annihilating to $W^+W^-$.} 
\label{fig:dens}
\vspace{-0.4cm}
\end{center}
\end{figure} 

The contours in Fig.~\ref{fig:dens} show the number density of suitable minihalos, and hence the maximum possible number of DM spikes, as a function of distance from the Solar System, $D$, and declination angle, $\delta$, for fixed values of right ascension, $\alpha$. $N_{\rm sp}^{(z_f)}(R)$ is defined as spherically symmetric function of galactic radius in Eq.~\ref{eq:Nsp}. The shapes of the contours in each panel are thus two dimensional projections of $N_{\rm sp}^{(z_f)}(R)$ transformed to coordinates centered around the Solar System. For example, the top panels show number densities of suitable minihalos potentially detectable from Earth for right ascension fixed in the direction of galactic center. The peaks of the respective $N_{\rm sp}^{(z_f)}(R)$ distributions are located near galactic center, which is visible in the top panels around $\delta = -28.94^\circ$ and $D = 8.5 \, {\rm kpc}$. The middle and bottom panels similarly show contours of $N_{\rm sp}^{(z_f)}(R)$ as functions of $\delta $ and $D$ for right ascension fixed perpendicular to and away from galactic center, respectively.

We only show the density of DM spikes in the galaxy, $N_{\rm sp}^{(z_f)}$, for the Late and Early termination of Pop III.1 star formation, with the density of spikes in the galaxy for the Intermediate $z_f$ scenario being similar to the former. The later star formation is allowed to precede, the more minihalos become massive enough to host Pop III.1 stars even though the halo mass threshold in Eq.~\ref{eqn:mmin} increases with time. More specifically, we can see the number of DM spikes in the Early $z_f$ scenario is suppressed by over an order of magnitude relative to the Late $z_f$ scenario. Also, the Early truncation of Pop III.1 star formation leads to a distribution of DM spikes which falls off more steeply away from galactic center. Thus, as we will demonstrate below, significantly fewer DM spikes could be detected as point sources (PSs) in the Early $z_f$ scenario, and the associated limits on DM annihilation are also somewhat less robust to any potential disruption of the spikes located closer to the galactic center.

In order to compute annihilation signals from the distribution of DM spikes in the galaxy, we also must implement a model for the density profiles of the individual DM spikes. We estimate the adiabatic contraction of each minihalo using the Blumenthal et al. prescription~\cite{Blumenthal:1986}, which is based on the assumption that the orbital time of the particles is much longer than the infall time and, thus, only circular orbits need be considered. With the additional assumptions of spherical symmetry and conservation of angular momentum in the Blumenthal et al. method, the homologous contraction of the minihalos yields a relation between the initial and final DM and baryon masses contained within a spheres of given radii from the centers of the minihalos 
\begin{equation}
    \left[ M_{DM} (r) + M_{b} (r) \right] r = \left[ M_{DM} (r) + M_{b} (r_f) \right] r_f
     \, .
\end{equation}
Following previous work~\cite{Gnedin:2004cx,Iocco:2008rb,Freese:2008hb,Bertone:2005xz,Bertone:2009kj,Sandick:2010yd,Sandick:2011zs}, we assume that the initial mass profiles, $M_{DM} (r)$ and $M_{b} (r)$, can be approximated as NFW~\cite{Navarro:1995iw} and that 15\% of the total mass in each minihalo is baryonic. The final DM distribution can then be estimated given the final baryon mass profile, $M_{b} (r_f)$, from numerical simulations of Pop III.1 star formation. In particular, simulations in Ref.~\cite{Abel:2001pr} showed the hydrogen number density relevant for collapsing protostellar gas in minihalos has a cored profile with a relatively constant $n \sim 10^{13} \, {\rm cm}^{-3}$ out to radii of $\sim 10^{14} \, {\rm cm}$ and falls off as $n \propto r^{-2.3}$ at larger radii. 

Several groups have verified that the density of the DM spikes using this simple Blumenthal et al. method is accurate to within a factor of two when compared to numerical simulations of DSs~\cite{Iocco:2008rb,Natarajan:2008db}. Ref.~\cite{Freese:2008hb} compared density profiles of adiabatically contracted minihalos based on the Blumenthal et al. prescription to a modified method derived by Young~\cite{Young:1980}, which takes into account the adiabatic invariance of the radial action in addition to that of the angular momentum, and found the respective DM spike profiles to be consistent within a factor of two.
 
The contracted DM halo profiles today can be treated in the following way. At radii far away from the central BH, the density profile falls off as a power law independent of DM mass. In the central regions, closest to the BH, some of the DM has annihilated away in the time since the formation of the central mass. Following Ref.~\cite{Bertone:2005xz}, we account for the annihilation of DM over the lifetime of the spike, $t_{BH}$, by imposing a high density cut-off at small radii:
\begin{equation} 
\rho_{\rm max}= \frac{m_{\chi}}{\langle \sigma v  \rangle  t_{BH}} \, \label{eqn:cutoff}
 \end{equation}
where $m_{\chi}$ is the WIMP mass, $\langle \sigma v  \rangle$ is the annihilation cross section, and $t_{BH} \sim 1.3 \times 10^{10}$ years for a Pop III.1 star formed at $z=15$. As described in \cite{Sandick:2010yd}, the resulting DM spike profiles for $10-10^5M_\odot$ BHs hosted by minihalos of mass $\sim 10^6 M_\odot$ have a constant mass density of $\rho_{\rm max} \sim 10^9 \, {\rm GeV / cm^3}$ out to a radius of $ \sim 10^{16} \,$cm assuming $m_{\chi} = 100 \,$GeV and $\langle \sigma v  \rangle=3 \times 10^{-26}$ cm$^3$ s$^{-1}$. For more massive BHs, adiabatic contraction yields higher density DM spikes. Larger BH masses therefore lead to stronger WIMP annihilation signals, as well as $\rho_{\rm max}$ being reached at larger radii. While the precise radius associated with the high density cut-off depends on the parameters in Eq.~\ref{eqn:cutoff} as well as the BH mass, the radial dependence of the spike density profiles at larger radii all follow similar power laws associated with the adiabatic contraction of the initial NFW profile.\footnote{In this work, we assume that DM in not self-interacting. For recent studies of DM spike density profiles formed under the assumption of self-interacting DM, see for example Refs.~\cite{Shapiro:2014oha,Feng:2021qkj}.}

It is important to note that individual DM spikes forming earlier will yield more dense profiles. Prior to the host minihalos contracting, the DM densities are diluted as the Universe expands. Thus, everything else being equal, the signal from WIMP annihilation in a single DM spike formed earlier will be larger than for a spike formed at later times. However, when considered in combination with the relative suppression of $N_{\rm sp}^{(20)}$ associated with earlier termination of Pop III.1 star formation, the constraints on the Early $z_f$ scenario are still considerably weaker than the Intermediate or Late $z_f$ scenarios. Independent of the DM spike distributions in the galaxy, we can also see from Eq.~\ref{eqn:cutoff} that larger WIMP masses yield larger $\rho_{\rm max}$. The signal from WIMP annihilation in DM spikes is less sensitive to WIMP mass than ID searches which only consider smooth parts of galactic halo, as the DM annihilation rate described in the next section is $\propto m_{\chi}^{-9/7}$~\cite{Bertone:2009kj} rather than the characteristic $\propto m_{\chi}^{-2}$. The more mild dependence of the WIMP annihilation signal on WIMP mass is particularly beneficial for the sensitivity of neutrino experiments to DM spikes because AN and IC are much more efficient at detecting the higher energy neutrinos relevant for the signal from higher mass WIMPs.
 
\section{Neutrino point sources} 
\label{sec:signal}
 
In this section, we calculate the flux of neutrinos observed at Earth arising from WIMP annihilation in a single DM spike and the corresponding number of events predicted in AN and IC neutrino PS searches. The WIMP annihilation rate in a spike with DM density $\rho_{DM}$ is 
\begin{equation} \Gamma = \frac{\langle \sigma v  \rangle}{2m_{\chi}^2}  \int_{r_{\rm min}}^{r_{\rm max}} dr \, 4\pi r^2 \rho^2_{DM} \, .
\label{eqn:annRate}
\end{equation}
The DM density profile (squared) of each spike is integrated from $r_{\rm min}$, taken as four times the Schwarzchild radius of the associated BH, to $r_{\rm max}$. For the DM density profiles described in the previous section, the corresponding WIMP annihilation rates are largely unchanged so long as $r_{\rm min} \lesssim 10^{14} \,$cm and $r_{\rm max} \gtrsim 10^{17} \,$cm. Similar to the relatively mild dependence of the WIMP annihilation rate on $m_{\chi}$, the cutoff of the DM density within the spike described by Eq.~\ref{eqn:cutoff} yields a signal $\propto \langle \sigma v  \rangle^{2/7}$~\cite{Bertone:2009kj} rather than the more characteristic linear dependence on the WIMP annihilation cross section associated with indirect detection searches. For simplicity, we take the annihilation cross section to be that of a thermal WIMP which comprises the measured DM density in the Universe, $\langle \sigma v  \rangle=3 \times 10^{-26}$ cm$^3$ s$^{-1}$, and study benchmark WIMP mass values of $m_\chi = 500 \, {\rm GeV}$ and $m_\chi = 5 \, {\rm TeV}$.

We assume that DM annihilates to $b\bar b$, $W^+W^-$, $\tau^+\tau^-$ or $\mu^+\mu^-$ final states. We use the PPPC4DMID~\cite{Cirelli:2010xx,Ciafaloni:2010ti} neutrino spectra, $dN_{f \to \nu_\ell}/dE_\nu$, for neutrino flavors $\ell=\{ e, \mu, \tau \}$ produced in WIMP annihilation to each final state $f$. Although some combination of final states is possible for any particular WIMP model, for simplicity we analyze each final state independently and assume that the associated branching fraction for WIMP annihilation is 1. After taking into account the probability for $\nu_\ell$ to oscillate into $\nu_{\ell'}$ in vacuum, $P(\nu_\ell \to \nu_{\ell'})$, we can write the differential flux of neutrinos from an individual DM spike as
\bea \label{eqn:spikeFlux}
{ d \Phi_{f \to \nu_{\ell'}} \over d E_\nu} = {\Gamma \over 4 \pi D^2} \sum_{\ell = e, \mu, \tau} P\left( \nu_\ell \to \nu_{\ell'} \right) {d N_{f \to \nu_\ell} \over d E_\nu} \, ,
\eea
where the distance from the solar system to the DM spike, $D$, is assumed to be much larger than the spatial extent of the spike.\footnote{For DM spikes closer to the solar system, the integral in Eq.~\ref{eqn:annRate} must be calculated of along the line of sight towards the spike and over the solid angle of interest. Note that for all DM spikes located at distances far enough away not to be detected as PSs by AN or IC in our analysis, the annihilation rate in each spike is well described by Eq.~\ref{eqn:annRate}.} We also assume the DM spikes are sufficiently far from the solar system such that the vacuum oscillations can be calculated in the long baseline limit with the oscillation probabilities of interest approximated by $P\left( \nu_\ell \to \nu_{e} \right) = (0.6,0.2,0.2)$ and $P\left( \nu_\ell \to \nu_{\mu} \right) = (0.2,0.4,0.4)$. 

In the remainder of this section, we calculate the minimum distance from our solar system, $D_{\rm min}$, in various regions of the sky, for a single DM spike to be located in order to not be detected as a PS by AN or IC. In addition, we calculate the associated diffuse neutrino flux from all DM spikes at distances from the solar system larger than $D_{\rm min}$ and compare to the relevant measurements of the atmospheric neutrino flux by AN and IC. For all cases investigated in this work, the most stringent limits on Pop III.1 star formation are from PS neutrino searches, and the diffuse neutrino flux from DM spikes is far below the observed atmospheric neutrino flux in nearly all cases.

AN and IC provide effective areas for the detection of neutrino PSs averaged over different parts of the sky. For AN, the three regions of the sky relevant for our analysis are given by the declination ranges $\left[-90^\circ ,-45^\circ \right] \,$ (referred to as AN1), $\left[-45^\circ ,0^\circ \right] \,$ (AN2) and $\left[0^\circ ,45^\circ \right] \,$ (AN3). While the IC PS search covers the full sky, we focus on the regions where the analysis has sensitivity to neutrinos with energies $E_\nu < 10 \, {\rm TeV}$. Also, the IC limits on neutrino PSs arise from a combined analysis using data from a variety of detector configurations, with each analysis providing effective areas averaged over different declination ranges. Thus, we define the IC declination ranges for our analysis from the overlap of the ranges in each individual IC analysis. The appropriate declination ranges are $\left[-30^\circ ,-5^\circ \right] \,$ (IC1), $\left[-5^\circ ,0^\circ \right] \,$ (IC2), $\left[0^\circ ,30^\circ \right] \,$ (IC3), $\left[30^\circ ,60^\circ \right] \,$ (IC4) and $\left[60^\circ ,90^\circ \right] \,$ (IC5). 

\begin{table}
   \centering
   \begin{tabular}{c c c c c c}
      \hline\hline
      EX$i$ & $\delta$ Range & $<$ Flux (${\rm GeV} \, {\rm cm}^{-2} {\rm s}^{-1}$) & EV & Acceptance (${\rm GeV}^{-1} {\rm cm}^2 \, {\rm s}$) & $ < N_{\rm EV}^{{\rm EX}i}$ \\
      \hline
      AN1 & $\left[-90^\circ ,-45^\circ \right] $ & 
      $6.3 \times 10^{-9}$ & TR & $2.8 \times 10^8$ & $1.8$  \\
      \, & \, & \, & SH & $7.1 \times 10^7$ & $0.4$  \\
            \hline
      AN2 & $\left[-45^\circ ,0^\circ \right] $ &
      $8.3 \times 10^{-9}$ & TR & $2.0 \times 10^8$ & $1.7$  \\
      \, & \, & \, & SH & $5.8 \times 10^7$ & $0.5$  \\
            \hline
      AN3 & $\left[0^\circ ,45^\circ \right] $ &
      $1.2 \times 10^{-8}$ & TR & $1.3 \times 10^8$ & $1.5$  \\
      \, & \, & \, & SH & $4.5 \times 10^7$ & $0.5$  \\
      \hline
      IC1 & $\left[-30^\circ ,-5^\circ \right] $ &
      $1.3 \times 10^{-9}$ & TR & $4.0 \times 10^9$ & $5.4$  \\
            \hline
      IC2 & $\left[-5^\circ ,0^\circ \right] $ &
      $2.6 \times 10^{-10}$ & TR & $1.5 \times 10^{10}$ & $3.9$  \\
            \hline
      IC3 & $\left[0^\circ ,30^\circ \right] $ &
      $3.1 \times 10^{-10}$ & TR & $1.9 \times 10^{10}$ & $5.8$  \\
            \hline
      IC4 & $\left[30^\circ ,60^\circ \right] $ &
      $4.5 \times 10^{-10}$ & TR & $1.4 \times 10^{10}$ & $6.3$ \\
      \hline 
      IC5 & $\left[60^\circ ,90^\circ \right] $ &
      $9.9 \times 10^{-10}$ & TR & $1.4 \times 10^{10}$ & $14$ \\
      \hline \hline
   \end{tabular}
   \caption{Summary of point source limits from ANTARES (AN)~\cite{ANTARES:2017dda} and IceCube (IC)~\cite{IceCube:2019cia} averaged over various declination regions. The third column shows the averaged limit on the neutrino flux normalization assuming a $\propto E_\nu^{-2}$ source spectrum. The corresponding average acceptance for track (TR)- or shower (SH)-like events is shown in the fifth column, while the sixth column shows the derived limit on the number of events which can be used to constrain the neutrino flux from dark matter spikes.}
   \label{tab:data}
\end{table}

Given the neutrino flux from an individual DM spike calculated using Eq.~\ref{eqn:spikeFlux}, the number of track (TR)- or shower (SH)-like events (denoted by ${{\rm EV} = \{{\rm TR},{\rm SH} }\}$) expected in each of the $i$th declination ranges (defined in the previous paragraph and Table~\ref{tab:data}) at either experiment (denoted by ${{\rm EX}i = \{{\rm AN}i,{\rm IC}i }\}$) is given by
\bea
N_{\rm EV}^{{\rm EX}i} = 
\sum_{j \in {\rm EX}i} T_{\rm exp}^{j}
\sum_{\ell' \in {\rm EV}} \int_{E_{\rm th}}^{m_\chi} d E_\nu 
{ d \Phi_{f \to \nu_{\ell'}} \over d E_\nu} A_{ \ell' {\rm EV}  }^{j} \left(E_\nu \right) \, ,
\label{eqn:NEV}
\eea
where the neutrino energy threshold is $ E_{\rm th} =100 \, {\rm GeV}$ for both experiments. The sum over neutrino flavors, $ \ell' $, runs over the relevant contributions to each event type. As described in the previous paragraph, the IC neutrino point source limits arise from a combination of individual analyses corresponding to several different detector configurations. In Eq.~\ref{eqn:NEV}, the $j$th contribution from each individual analysis to the number of events in a given sky region is characterized by an exposure time $T_{\rm exp}^{j}$ and an effective area $A_{ \ell' {\rm EV}  }^{j}$, which is specified for the flavor of neutrino contributing to either TR or SH events. In both AN and IC, the events which are most relevant to setting limits on the neutrino flux from a single DM spike arise from charged current $\nu_\mu$ interactions that yield TR events. For AN, we also include SH events which are associated with charged current $\nu_e$ and neutral current $\nu_\mu$ interactions. 

For each experiment, limits on the normalization of the neutrino flux from a single PS are presented as a function of declination angle assuming a source spectrum with an energy dependence $\propto E_\nu^{-2}$. The associated limits on the number of events as a function of declination angle can be calculated by multiplying the limits on the flux normalization by the acceptance of each experiment to TR or SH events. To approximate limits on $N_{\rm EV}^{{\rm AN}i}$, we average the limits on the flux normalization and the acceptances over the declination ranges corresponding to the averaged effective areas provided in the AN analysis. For the analogous limits on $N_{\rm TR}^{{\rm IC}i}$, we calculate the averaged acceptance explicitly in each sky region by taking the sum of acceptances from the relevant declination ranges of each individual analysis. The $j$th contribution to the acceptance from each individual analysis is given by the product of the exposure, $T_{\rm exp}^j$, with the convolution of the respective effective area and an $\propto E_\nu^{-2}$ source spectrum.\footnote{We have also checked that an analogous calculation of the averaged AN acceptances are consistent with the acceptances which are directly provided.} The relevant quantities derived from AN~\cite{ANTARES:2017dda} and IC~\cite{IceCube:2019cia} data are summarized in Table~\ref{tab:data}. 

\begin{figure}
\begin{center}
\includegraphics[width=3in]{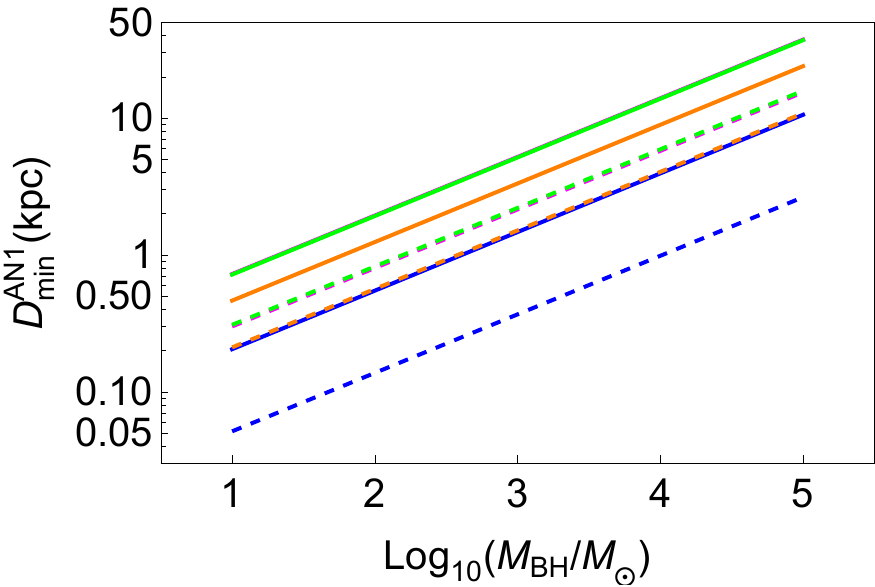}\hfill
\includegraphics[width=3in]{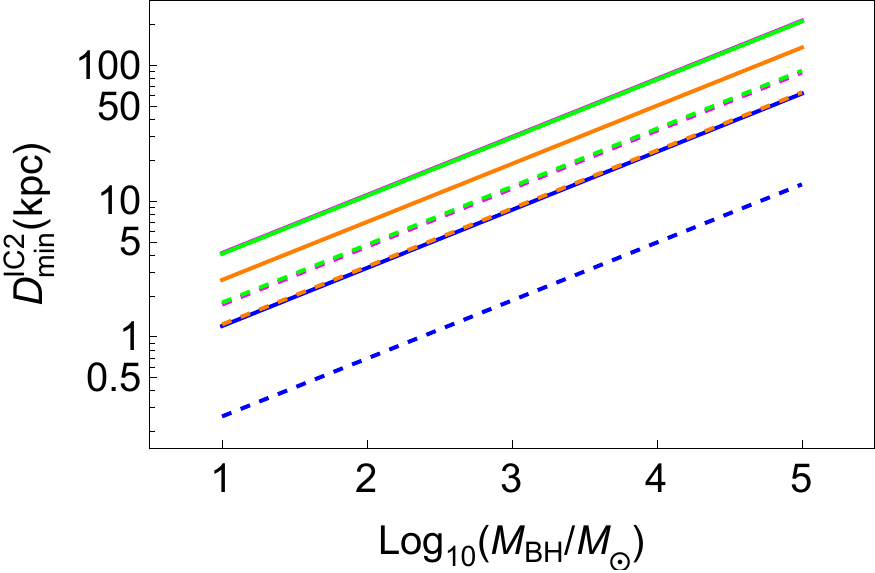}
\includegraphics[width=3in]{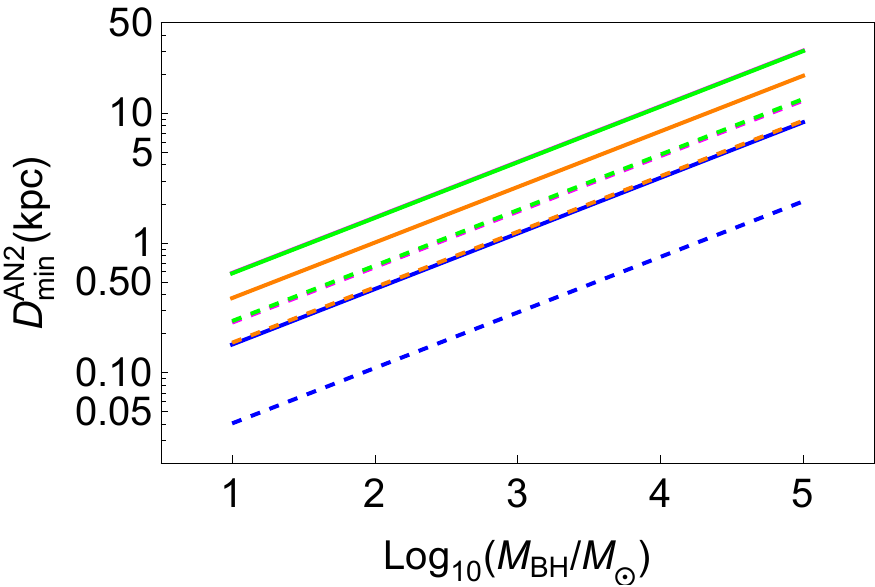}\hfill
\includegraphics[width=3in]{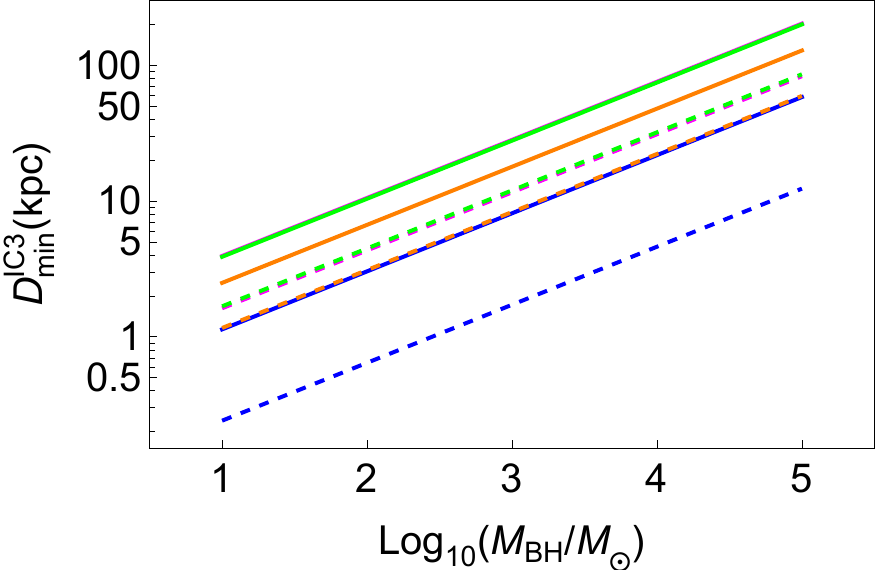}
\includegraphics[width=3in]{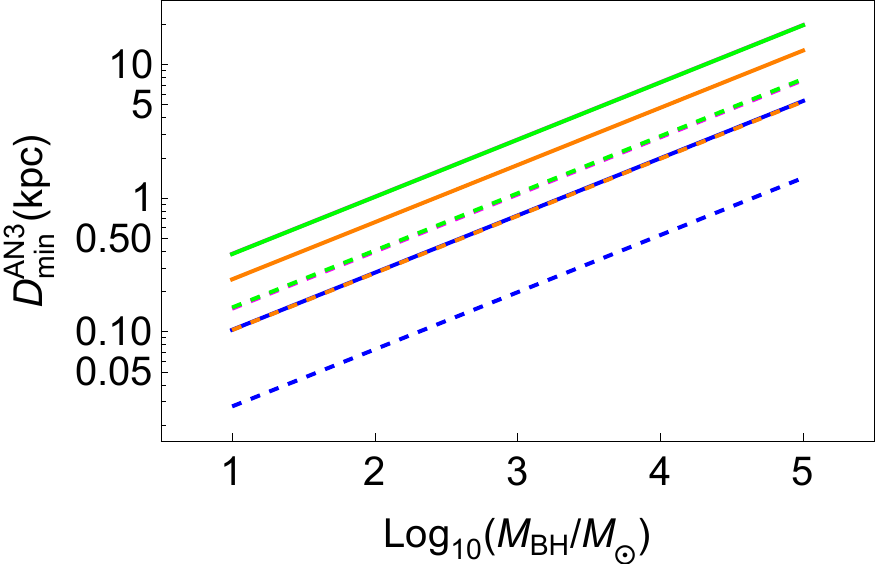}\hfill
\includegraphics[width=3in]{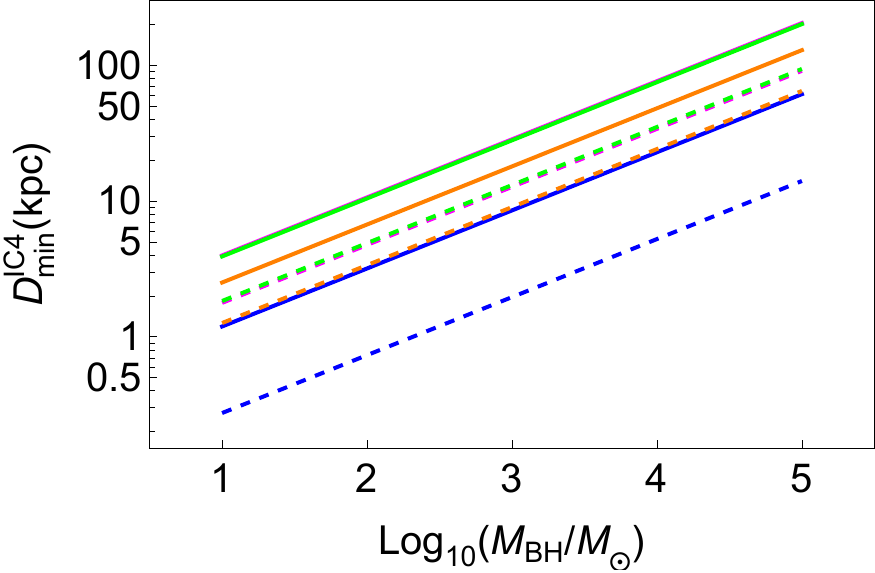}
\includegraphics[width=3in]{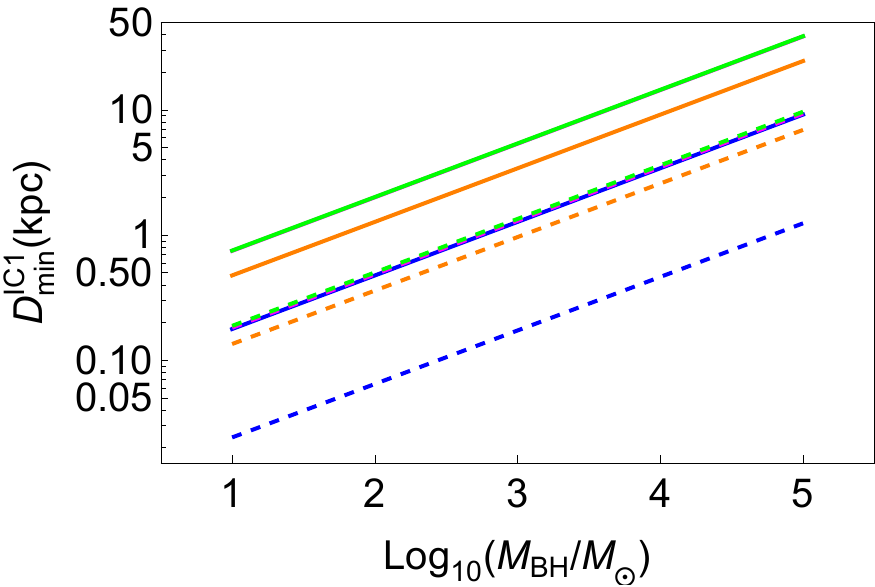}\hfill
\includegraphics[width=3in]{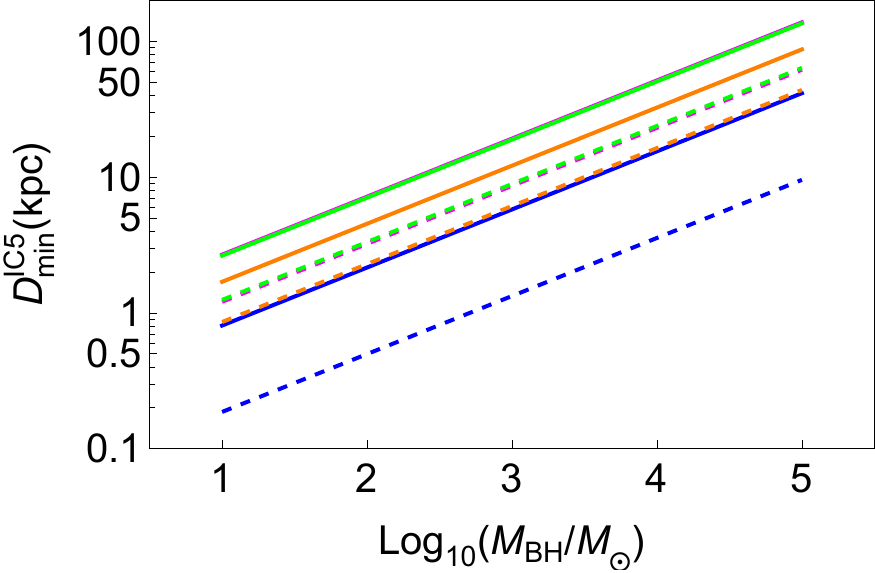}
\vspace{-0.0cm}
\caption{The minimum distance away from the solar system, $D_{\rm min}$,  for an individual dark matter spike to have avoided detection in neutrino point source searches at ANTARES (AN) and IceCube (IC) over declination ranges (left, from top to bottom) $\left[-90^\circ ,-45^\circ \right] \,$, $\left[-45^\circ ,0^\circ \right] \,$ $\left[0^\circ ,45^\circ \right] \,$, $\left[-30^\circ ,-5^\circ \right] \,$  and (right, from top to bottom) $\left[-5^\circ ,0^\circ \right] \,$, $\left[0^\circ ,30^\circ \right] \,$ $\left[30^\circ ,60^\circ \right] \,$, $\left[60^\circ ,90^\circ \right] \,$. The $D_{\rm min} \,$ in each panel are shown for $m_\chi = 500 \,$GeV (dashed) and $m_\chi = 5 \,$TeV (solid) for WIMP annihilation to $b\bar b$ (blue), $W^+W^-$ (orange), $\tau^+\tau^-$ (magenta) and $\mu^+\mu^-$ (green).} 
\label{fig:dmin}
\vspace{-0.4cm}
\end{center}
\end{figure} 

For each of the respective AN and IC sky regions, we calculate $D_{\rm min}$ for a variety of WIMP annihilation channels with $m_\chi = 500 \, {\rm GeV}$ and $m_\chi = 5 \, {\rm TeV}$. For each WIMP mass and annihilation channel, we show $D_{\rm min}$ for BH masses from $10 \, M_\odot$ to $10^5 M_\odot$ assuming $z_f = 15$ in Fig.~\ref{fig:dmin}. In all cases, we see that $D_{\rm min}$ is maximized for $M_{BH}= 10^5 M_\odot$ due to the enhanced DM density profiles associated with larger BH masses. For each sky region, we can see that $D_{\rm min}$ increases for larger WIMP masses since the AN and IC analyses are more sensitive to higher energy neutrinos. $D_{\rm min}$ also increases for leptonic WIMP annihilation channels which produce neutrinos more efficiently. Due to the enhanced sensitivity of the IC analysis to PSs located below the horizon, the $D_{\rm min}$ for the IC sky regions with $\delta \gtrsim -5^\circ $ (i.e. IC2, IC3, IC4, IC5) can be up to an order of magnitude larger compared to the that of other regions (i.e. AN1, AN2, AN3, IC1). While the parameter dependence of $D_{\rm min}$ shown in Fig.~\ref{fig:dmin} remains qualitatively the same under different assumptions for the termination of Pop III.1 star formation, as described in Sec.~\ref{sec:spikes}, the DM density is marginally enhanced (suppressed) for spikes formed assuming $z_f = 20$ ($z_f = 10$) compared to $z_f = 15$. 

The effects of spikes with relatively enhanced or suppressed DM density are demonstrated by the horizontal black lines in Fig.~\ref{fig:dens}, which show the largest $D_{\rm min}$ as a function of $\delta$ for $m_\chi = 500 \, {\rm GeV}$ and $m_\chi = 5 \, {\rm TeV}$ WIMPs annihilating to $W^+W^-$ assuming either $z_f = 10$ or $z_f = 20$. For this representative WIMP benchmark, we see the early termination of Pop III.1 star formation results in a $\sim 20 \% \,$ larger $D_{\rm min}$ across the full range of declination angles when compared to the Late $z_f$ scenario. However, the relative enhancement of $N_{\rm sp}^{(10)}$ more than compensates for the relative suppression of the signal from the individual DM spikes. As we discuss in Sec.~\ref{sec:constraints}, the fraction of subhalos which host DM spikes can be constrained by integrating the number density of spikes, $N_{\rm sp}^{(z_f)}$, over the volume contained within the ``$D_{\rm min}$ surface'' defined as a function of $\delta$ across the sky. Comparing $N_{\rm sp}^{(10)}$ to $N_{\rm sp}^{(20)}$ for $D < D_{\rm min}$ in Fig.~\ref{fig:dens}, we see that the number density of DM spikes in the galaxy can be up to an order of magnitude larger for $z_f = 10$ when compared to $z_f = 20$. It is also worth noting that, for nearly all cases we consider, the volume contained by the $D_{\rm min}$ surface does not include many of the DM spikes which are concentrated in the galactic center. In principle, the exclusion of these DM spikes limits the sensitivity of neutrino PS searches but, as we discuss further in Sec.~\ref{sec:constraints}, spikes formed near the galactic center are more likely to have been disrupted.
 
\section{Constraints on Pop III.1 star formation} \label{sec:constraints}

For the calculation of the neutrino signal from DM spikes in the previous section, we assume that all viable minihalos have hosted Pop III.1 stars that subsequently collapsed to BHs which have not undergone mergers. While this assumption yields the largest possible neutrino signal from WIMP annihilation in DM spikes, the lack of neutrino PSs observed by AN and IC for all of the cases we have considered implies that either not all viable minihalos hosted Pop III.1 stars, or, for minihalos that have, the associated DM spikes have undergone mergers which have erased the overdensities of DM. We parameterize the fraction of viable minihalos that eventually contribute to the neutrino signal via their associated DM spikes as $f_{\rm DS}$. For each case we consider, we require the number of DM spikes within the volume contained by the $D_{\rm min}$ surface to be less than $1$ such that
\begin{equation}
    f_{\rm DS} \int_{s<D_{\rm min}}d^3 \mathbf{s} \, N_{\rm sp}^{(z_f)}(R) < 1 \, ,
\end{equation}
where $s$ is the distance from the solar system and $R$ is the radial distance from galactic center. Note that the angular dependence of the volume integral over the DM spike density is non-trivial due to the typically anisotropic shape of the $D_{\rm min}$ surface, as shown in Fig.~\ref{fig:dens}.

\begin{figure}
\begin{center}
\includegraphics[width=3in]{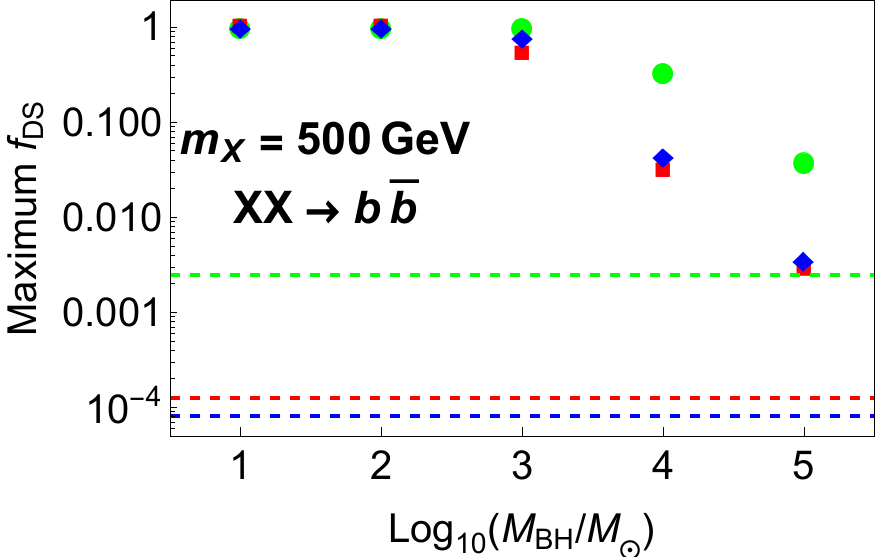}\hfill
\includegraphics[width=3in]{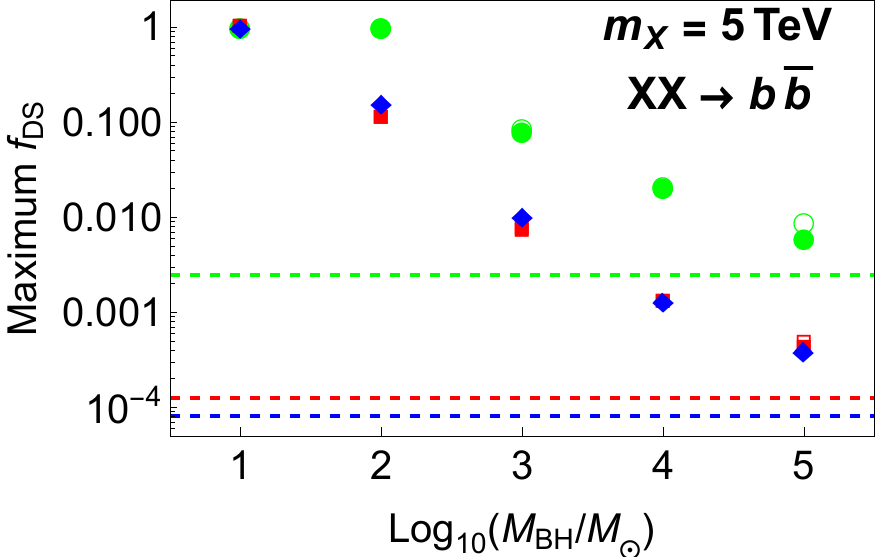}
\includegraphics[width=3in]{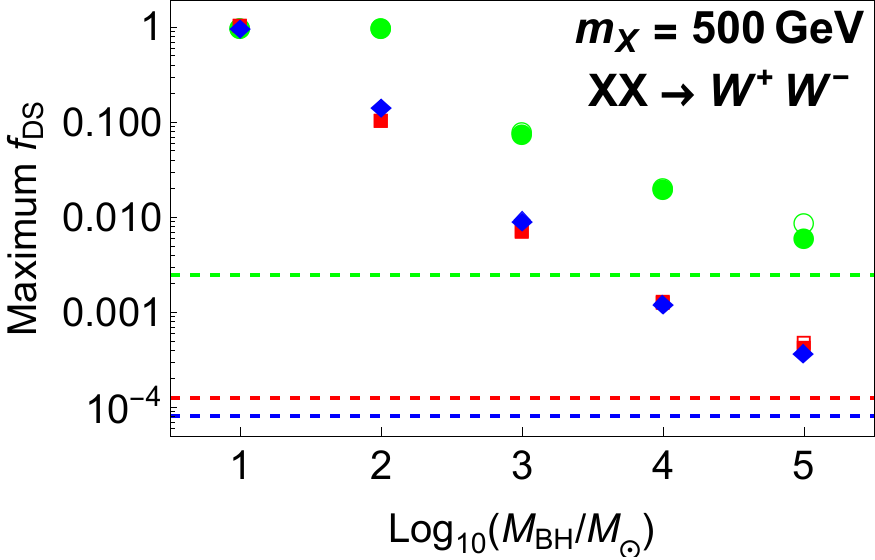}\hfill
\includegraphics[width=3in]{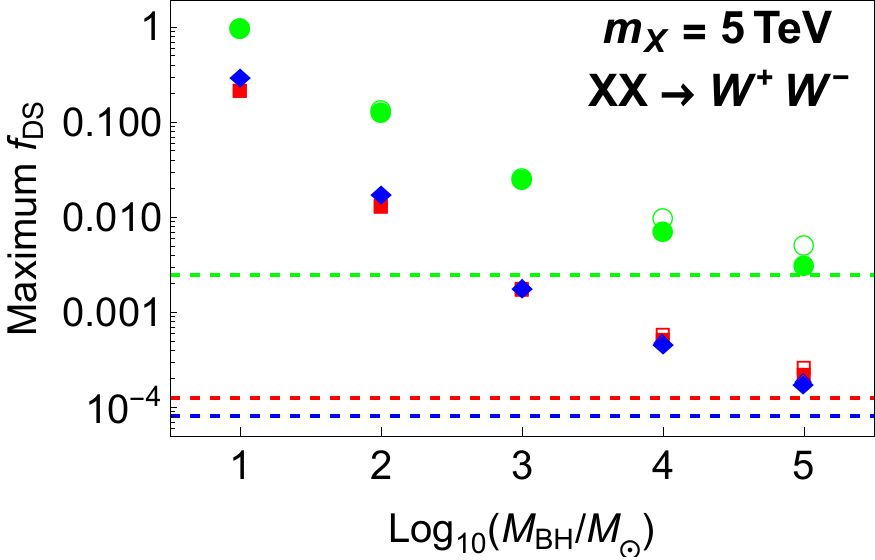}
\includegraphics[width=3in]{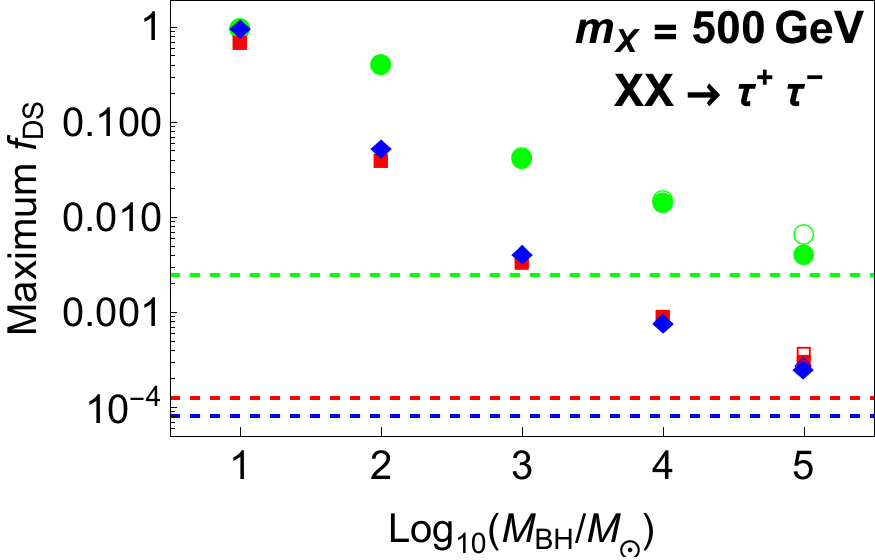}\hfill
\includegraphics[width=3in]{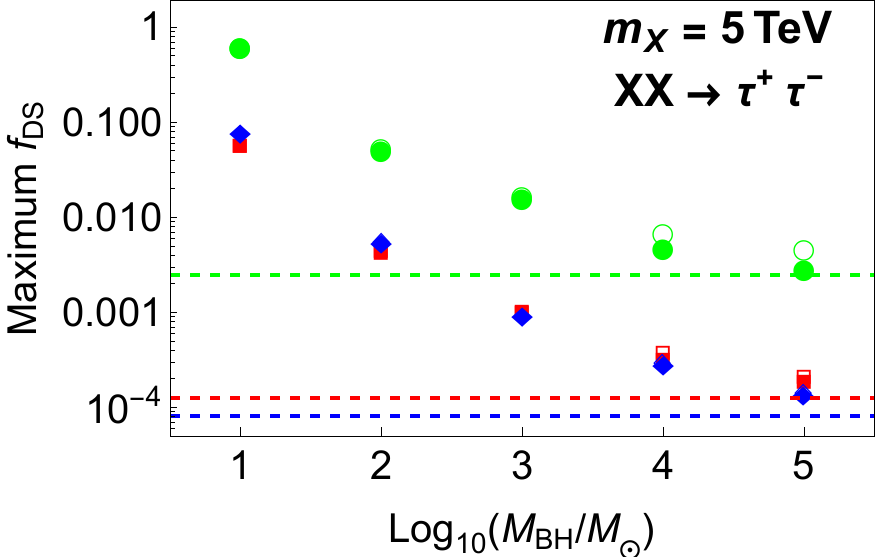}
\includegraphics[width=3in]{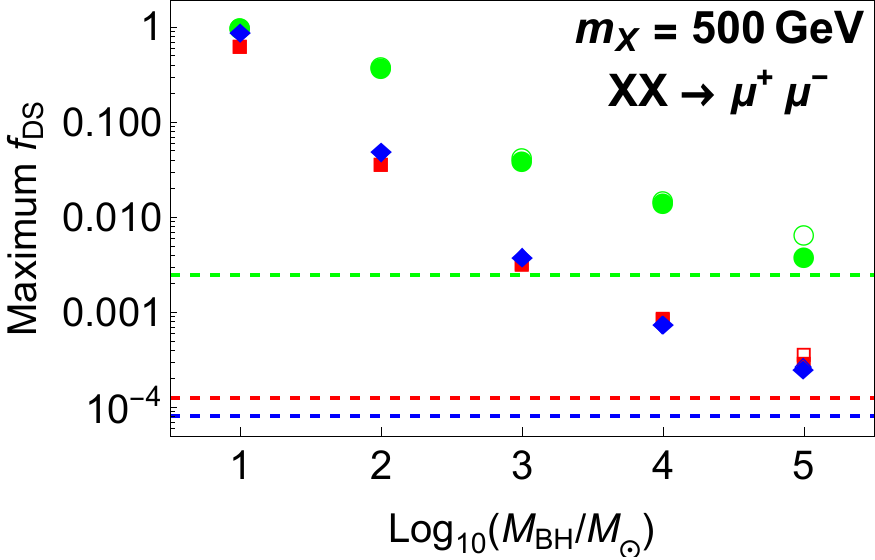}\hfill
\includegraphics[width=3in]{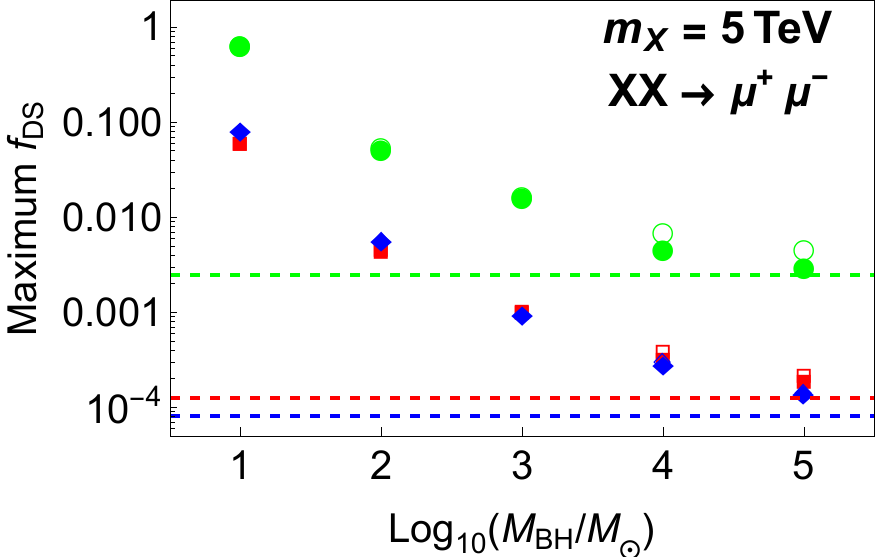}
\vspace{-0.0cm}
\caption{Maximum fraction of minihalos, $f_{DS}$, consistent with neutrino point source limits which can potentially host dark matter spikes as a function of central BH mass for $m_\chi = 500 \,$GeV (left) and $m_\chi = 5 \,$TeV (right) WIMP annihilation into $b \bar b$, $W^+W^-$, $\tau^+\tau^-$ and $\mu^+\mu^-$. Green circles, red squares, and blue diamonds are for Early, Intermediate, and Late $z_f$ scenarios, respectively. The solid markers are the limits accounting for the entire spike distribution in the galaxy, while the open markers are the limits excluding spikes from within $5 \,$kpc of the Galactic center. The horizontal dashed lines show the $f_{DS}$ corresponding to a single DM spike for Early (green), Intermediate (red) and Late (blue) $z_f$ scenarios.}
\label{fig:combPSfDS}
\vspace{-0.0cm}
\end{center}
\end{figure}

The associated upper limits on $f_{\rm DS}$ are shown for $m_\chi = 500 \,$GeV and $m_\chi = 5 \,$TeV in Fig.~\ref{fig:combPSfDS}. We consider WIMP annihilation to $b\bar b$, $W^+W^-$, $\tau^+\tau^-$ and $\mu^+\mu^-$ final states, while assuming either Early, Intermediate, or Late $z_f$ scenarios for the termination of Pop III.1 star formation. We plot the limits on $f_{\rm DS}$ as a function of BH masses $10 M_{\rm \odot}$, $10^2 M_{\rm \odot} $, $10^3 M_{\rm \odot}$, $10^4 M_{\rm \odot}$ and $10^5 M_{\rm \odot}$. As expected from the parameter dependencies of the $D_{\rm min}$ surface described in the previous section, the cases with the strongest constraints on $f_{\rm DS}$ arise for heavier WIMPs forming spikes around the most massive BHs and annihilating through more leptonic channels. Also, we see that the bounds are up to about an order of magnitude stronger in the Intermediate and Late Pop III.1 star formation termination scenarios compared to the Early $z_f$ scenario. 

More specifically, for the least constrained WIMP annihilation channel, $b\bar b$, we see that PS searches from AN and IC are not sensitive to signals from DM spikes forming around $10 M_{\rm \odot}$ BHs. The constraints are more stringent for heavier BHs, such that $f_{\rm DS} \lesssim 10 \%$ for $10^5 M_{\rm \odot}$ BHs for all $z_f$ scenarios. The most stringent constraints arise for WIMPs annihilating to a $\mu^+\mu^-$ final state around $\gtrsim 10^3 M_{\rm \odot}$ BHs forming in the Intermediate and Late $z_f$ scenarios, with $f_{\rm DS} \lesssim 0.1 \%$. For the Early scenario in this case, the constraints on $f_{\rm DS}$ are weaker but approach the limit where no viable minihalos can host DM spikes for $10^5 M_{\rm \odot}$ BHs. Regardless of annihilation channel or $z_f$, we see that $f_{\rm DS} \lesssim 10 \%$ for $m_\chi = 5 \,$TeV WIMPs annihilating in DM spikes formed around $\gtrsim 10^3 M_{\rm \odot}$ BHs.

These limits are complimentary to those arising from the gamma-ray signals from DM spikes presented in Refs.~\cite{Sandick:2010yd,Sandick:2011zs}. In particular, gamma-ray constraints are better able to constrain lighter WIMPs annihilating to hadronic final states, including $b\bar b$. However, for all other annihilation channels considered in this paper, PS neutrino searches constrain $f_{\rm DS}$ to be up to three orders of magnitude smaller than the analogous limits from PS and diffuse gamma-ray searches for $m_\chi = 500 \,$GeV and $ m_\chi = 5 \,$TeV. In contrast to the previous implementation of constraints from gamma-ray PS searches, not a single neutrino PS has been confirmed by either AN or IC. In gamma-ray searches, the signal flux from a single DM spike only need be less than the brightest associated or unassociated PS. The neutrino flux from the same DM spike must lie below the respective sensitivities of the AN and IC PS searches, resulting in comparatively larger $D_{\rm min}$ and more stringent constraints on $f_{\rm DS}$ for most cases.

In the limits discussed above, we have not accounted for the potential impact of tidal disruption on the DM spikes near the galactic center, where the DM density is expected to be largest. While Ref.~\cite{Bertone:2005xz} has shown that there might be only small effects in certain cases, different assumptions can lead to the destruction of nearly all substructure within $\mathcal{O}(1) \,$kpc of  galactic center. Therefore, in Fig.~\ref{fig:combPSfDS} we also show the limits on $f_{\rm DS}$ when DM spikes within $5 \,$kpc of the galactic center are excluded from the spike distribution. The results do not differ significantly at all for the Intermediate and Late $z_f$ scenarios, while we can see that there is a marginal weakening of bounds when considering $\gtrsim 10^4 M_{\rm \odot}$ BHs formed assuming the Early termination of Pop III.1 star formation. As shown in Fig.~\ref{fig:dens}, the constraints on $f_{\rm DS}$ from neutrino PSs are robust to the exclusion of DM spikes within $5 \,$kpc of galactic center since $D_{\rm min}$ in the direction of galactic center is typically smaller than the distance from the solar system to the galactic center, $\sim 8.5 \,$kpc. In Fig.~\ref{fig:dmin}, we can see that $D_{\rm min}^{\rm AN2}$ and $D_{\rm min}^{\rm IC1}$ only become slightly larger than the distance to galactic center for a subset of cases with $\gtrsim 10^4 M_{\rm \odot}$ BHs.

The constraints on Pop III.1 star formation described in this section must also be distinguished from the depletion of DM spikes by potential mergers between the associated central BHs. Ref.~\cite{Sandick:2010yd} demonstrates that the fraction of BHs which undergo mergers is largest for the most massive BHs formed in the cases with later termination of Pop III.1 star formation, in which the highest number of BHs can potentially be produced. For $f_{\rm DS} = 1$ in the Late $z_f$ scenario, the fraction of BHs which would have merged is estimated to be 0.76\%, 24\%, and 49\% for BH masses of $10^3 M_{\rm \odot}$, $10^4 M_{\rm \odot}$ and $10^5 M_{\rm \odot}$, respectively. The merger fraction is itself a function of $f_{\rm DS}$ such that it vanishes if only a small fraction of minihalos are host to Pop III.1 stars which subsequently collapse to form BHs. For $\lesssim 10^3 M_{\rm \odot}$ BHs, the limits on $f_{\rm DS}$ from PS neutrino searches are weaker but are also clearly robust to the depletion of DM spikes by BH mergers. In the case of $10^4 M_{\rm \odot}$ BHs, the constraints on $f_{\rm DS}$ are only $\gtrsim 24\%$ for the Early $z_f$ scenario with $m_\chi = 500 \,$GeV WIMPs annihilating to $b\bar b$. For all other cases with $10^4 M_{\rm \odot}$ BHs and all cases with $10^5 M_{\rm \odot}$ BHs, $f_{\rm DS} \lesssim 1 \%$ regardless of $z_f$ and, thus, mergers are sufficiently rare to ensure that all minihalos which are able to host DM spikes are potentially detectable as neutrino PSs. In summary, we see that neutrino PS searches are a powerful and robust probe of Pop III.1 star formation under a variety of WIMP annihilation scenarios. 

\section{Conclusions} \label{sec:Con}

In addition to signatures of dark matter interacting with the SM at direct detection and collider experiments, we can also probe the nature of dark matter through the indirect detection of the SM particles produced in dark matter annihilation. Signals of dark matter annihilating in our galaxy can potentially be enhanced if dark matter spikes form around the black hole remnants that remain after the collapse of Pop III.1 stars within dark matter minihalos. The largely unconstrained history of Pop III.1 star formation, in particular the redshift $z_f$ after which such stars cease to form, impacts both the distribution of dark matter spikes in the galaxy and the density profiles of individual spikes. The dark matter annihilation signal from an individual spike can vary depending on the annihilation cross section, final state, and dark matter mass, as well as the density profile associated with a central black hole of a given mass. While previous studies have mainly focused on gamma-ray signals from WIMP annihilation, in this paper we investigate the sensitivity of neutrino point source searches to dark matter spikes. 

We study Early ($z_f=20$), Intermediate ($z_f=15$) and Late ($z_f = 10$) benchmark scenarios for the termination of Pop III.1 star formation. For  black holes that seed  dark matter spikes, we consider both a range of masses suggested by models of stellar evolution, $10-10^2M_\odot$, and more massive $10^3-10^5M_\odot$ black holes potentially relevant for a dark star phase of Pop III.1 star formation. We assume a thermal WIMP annihilation cross section for WIMP masses $m_\chi = 500 \,$GeV and $m_\chi = 5 \,$TeV and consider WIMP annihilation to $b \bar b$, $W^+W^-$, $\tau^+\tau^-$, and $\mu^+\mu^-$ final states. To estimate constraints from neutrino point source searches at ANTARES and IceCube on the signal from individual dark matter spikes, we recast the constraints on the neutrino flux normalization assuming a $\propto E_\nu^{-2}$ source spectrum to limits on the number of events originating from various regions across the sky.

For each experiment and the relevant sky regions, we calculate the minimum distance, $D_{\rm min}$, that a dark matter spike must be located from the solar system in order not to have been detected as a point source. Given a choice of $z_f$ scenario, black hole mass, WIMP annihilation channel, and WIMP mass, we combine the $D_{\rm min}$ from every region of the sky to constitute a ``$D_{\rm min}$ surface.'' To approximate the fraction of minihalos which can host dark matter spikes, $f_{\rm DS}$, we integrate the dark matter spike density over the volume bounded by the $D_{\rm min}$ surface and require that the number of spikes contained within the volume be less than 1. While a more sophisticated analysis could, in principal, yield more robust constraints on $f_{\rm DS}$, we consider our estimates to be indicative for a signal with $\mathcal{O}(1)$ uncertainties arising from the formation of dark matter spikes. We also demonstrate that our analysis is virtually insensitive to the effects of tidal disruption on dark matter spikes located near the galactic center and the potential for mergers between the associated black holes. 

The results of our analysis can be summarized by Fig.~\ref{fig:combPSfDS}. We see that $f_{\rm DS}$ is more stringently constrained for the Intermediate and Late $z_f$ scenarios relative to the Early $z_f$ scenario. Dark matter spikes formed around heavier central black hole masses are generally more constrained than spikes formed around lighter black holes. In contrast to gamma-ray searches for dark matter spikes, neutrino point source searches are particularly sensitive to models with heavier WIMPs and leptonic annihilation channels. Due to the enhanced sensitivity of neutrino point source searches to the higher energy neutrinos produced in the annihilation of more massive WIMPs, we see that $f_{\rm DS} \lesssim 10 \%$ for $m_\chi = 5 \,$TeV WIMPs annihilating in DM spikes formed around $\gtrsim 10^3 M_{\rm \odot}$ BHs. As this result holds across all annihilation channels and $z_f$ scenarios we consider, it suggests the possibility of an extended dark star phase in Pop III.1 star formation for $m_\chi \gtrsim 1 \,$TeV is severely constrained by neutrino point source searches. More generally, due to the relative brightness of gamma-ray point sources compared to the lack of neutrino point sources detected by ANTARES or IceCube, neutrino signals provide for a powerful constraint on dark matter spikes and Pop III.1 star formation. Such constraints will only improve with the  with construction of KM3Net~\cite{KM3NeT:2018wnd}, which will combine the visibility of dark matter spikes concentrated towards the galactic center provided by ANTARES with the larger exposure of IceCube.

\acknowledgments
The authors would like to thank P.~Ullio for useful discussions. K.~Freese gratefully acknowledges support from the Jeff and Gail Kodosky Endowed Chair in Physics at the University of Texas, Austin; the U.S. Department of Energy, Office of Science, Office of High Energy Physics program under Award Number DE-SC-0022021 at the University of Texas, Austin.  K.~Freese, I.~Galstyan, and P.~Stengel are grateful for support from the Vetenskapsr{\aa}det (Swedish Research Council) through contract number 638-2013-8993 at Stockholm University. The work of P.~Sandick is supported by NSF grant PHY-2014075. The work of P.~Stengel is partially supported by the research grant ``The Dark Universe: A Synergic Multi-messenger Approach" number 2017X7X85K under the program PRIN 2017 funded by the Ministero dell'Istruzione, Universit{\`a} e della Ricerca (MIUR), and by the ``Hidden" European ITN project (H2020-MSCA-ITN-2019//860881-HIDDeN).

\printbibliography
\end{document}